\documentstyle[12pt]{article}

\def\thebibliography#1{\section{References}\list
 {[\arabic{enumi}]}{\settowidth\labelwidth{[#1]}\leftmargin\labelwidth
 \advance\leftmargin\labelsep
 \usecounter{enumi}}
 \def\newblock{\hskip .11em plus .33em minus .07em}
 \sloppy\clubpenalty4000\widowpenalty4000
 \sfcode`\.=1000\relax}

\begin{document}
\begin{flushright}
CU-TP-799
\end{flushright}
\begin{center}
\begin{title}
\title{\Large \bf Limitations on using the operator product expansion at small values of \ 
x*}
\end{title}

\vskip 10pt
\begin{author}       
\author{A.H. Mueller\footnotetext{*This research is supported in 
part by the Department of Energy under GRANT DE-FG02-94ER40819.}\\
Department of Physics, Columbia University\\
New York, New York 10027, USA}
\end{author}
\end{center}

\vspace{1.5in}
\abstract

 Limits on the regions of $Q^2$ and \ x\ where the operator product expansion can
be safely used, at small values of \ x\ are given.  For a fixed large $Q^2$ there is an 
$x_0(Q^2)$
such that for Bjorken x-values below $x_0$ the operator product expansion breaks down 
with
significant nonperturbative corrections occurring in the leading twist coefficient and 
anomlous
dimension functions due to diffusion of gluons to small values of transverse momentum.

\section{Introduction}

\ \ \ \ \ For hard scattering processes involving two transverse momentum scales, one of
which is the hard scattering scale, factorization and the Dokshitzer, Gribov, Lipatov,
Altarelli, Parisi (DGLAP) equation[1-3] furnish the basis for a systematic description of 
the
dependence of cross sections on the hard scale.  For example, in deep inelastic lepton 
proton
scattering the two scales are the inverse size of the proton, proportional to the QCD
$\wedge$-parameter, and the virtuality,\ Q,\ of the photon exchanged between the lepton 
and
the proton.  Factorization is given by the operator product expansion which separates the 
hard
and soft scales while the DGLAP equation governs the $Q^2-$dependence.

For processes having only one (hard) transverse momentum scale the Balitsky, Fadin, 
Kuraev,
Lipatov(BFKL) equation[4,5] determines the center of mass energy dependence when the 
hard
transverse momentum scale is held fixed.  BFKL effects also contribute to two-scale 
processes
like deep inelastic scattering, and these effects are accounted for through a resummation 
of
certain higher order terms in the anomalous dimension and coefficient functions which 
occur in
the operator product expansion[6-12].

However, even for processes having only one transverse momentum scale the BFKL 
equation
ultimately breaks down, at very high energies,because gluons diffuse away from the hard 
scale and
reach transverse momenta of size $\wedge$ where perturbation theory no longer 
applies[5,13].  It
can be expected that a similar phenomenon will occur when BFKL effects are included in 
the DGLAP
equation through resummation.  For deep inelastic scattering at moderate values of \ x\ 
the
leading  twist terms in the operator product expansion represent the structure functions so 
long
as $Q/\wedge >> 1.$  However, if $Q/\wedge$ is fixed at some large value the operator 
expansion
breaks down for sufficiently small values of \ x\ because of diffusion, in both the 
anomalous
dimension and coefficient functions, into regions of transverse momentum on the order of
$\wedge.$  This $breakdown$ of the operator product expansion is not due to higher twist 
terms,
but rather to an inability to properly separate hard and soft scales at very small values of \
x.\   Of course within a region of \ x,\ say $x_0 < x < 1,$ one can always choose a 
$Q(x_0)$
such that if $Q > Q(x_0)$ then the operator product expansion applies.  As we shall see a
little later on (See (40)) this region is determined by $\ell n{Q(x_0)\over \wedge}\geq [{7
N_c\zeta(3)\over \pi b} \ell n\  \bar{x}/x]^{1/3},$  with $b = {11 N_c-2N_f\over 12\pi},$ 
when a
resummation of leading logarithmic terms is taken into account.
$\bar{x}$ is a fixed parameter, 
$\bar{x}
\approx 1/10$ perhaps, not determined in our analysis.

Our model of diffusion into the infrared is the fixed-coupling BFKL equation.  Because 
the
fixed-coupling BFKL equation underestimates diffusion into the infrared, and 
overestimates
diffusion into the ultraviolet, our result on the  region of applicability of the operator
product expansion, (40), should be taken with some care.  This uncertainty is implicit 
in (40) in that we are unable to specify the scale of the running coupling in D.  In the limit
for $\ell n {Q(x_0)\over \wedge}$ given just above we have taken $\alpha = 
\alpha(Q(x_0))$ in
(40).
\vskip 10pt
\section{BFKL evolution and its limitations}
\vskip 6pt

The BFKL equation naturally applies to high energy cross sections were there is only a 
single
transverse momentum scale,  Q,  if that scale is hard enough so that the running coupling,
$\alpha(Q),$ is small.  While there are a number of processes where BFKL evolution can 
be directly
measured, for our purposes the heavy onium-heavy onium total cross section is the 
simplest to
consider.  The scale \ Q\ is determined by the inverse heavy onium radius.  The cross 
section is
given[14], in leading logarithmic approximation, by 

$$\sigma(Y) = \int d^2x\int_0^1 dz \int d^2 x^\prime \int_0^1 dz^\prime
\Phi(x^\prime,z^\prime)f(x^\prime,Y,x) \Phi (x,z)\eqno(1)$$

\noindent where $\Phi(x,z)$ is the square of the light-cone wave function of the onium 
with \ x\
the transverse coordinate separation of the heavy quark and heavy antiquark while\   z\   
is
the longitudinal momentum fraction of the heavy quark.  $Y = \ell n\  s/M^2$ with\ s\ the 
center of
mass energy squared and \ M\ the mass of the onium.  The BFKL impact parameter 
amplitude,
$f(x^\prime,Y,x),$ is given in terms of the momentum space amplitude by

 $$f(x^\prime,Y,x) = \int {d^2k^\prime\over (k^\prime)^2}\  {d^2k\over k^2}\ 
e^{i\b{k}\cdot\b{x}-i\b{k}^\prime\cdot\b{x}^\prime} f(k^\prime,Y,k).\eqno(2)$$

\noindent Asymptotically,

$$f(x^\prime,Y,x) \sim 4\pi x x^\prime \alpha^2 {e^{(\alpha_P-1)Y}\over \sqrt{{7\over 
2}\alpha
N_c\zeta(3)Y}} exp\{-{\ell n^2 x^\prime/x\over 4 D Y}\}\eqno(3)$$

\noindent and

$$f(k^\prime,Y,k) \sim {\alpha^2\over \pi kk^\prime}\  {e^{(\alpha_P-1)Y}\over 
\sqrt{{7\over
2}\alpha N_c\zeta(3)Y}}\  exp\{-{\ell n^2\  k^\prime/k\over 4 D Y}\}\eqno(4)$$

\noindent where $\alpha_P-1={4\alpha N_c\over \pi}\ell n\ 2$\ and\   $D={7\alpha 
N_c\zeta(3)\over
2\pi}$.   Using (3) in (1) gives

$$\sigma \sim 16\pi R^2 \alpha^2 {e^{(\alpha_P-1)Y}\over \sqrt{{7\over 2}\alpha
N_c\zeta(3)Y}}\eqno(5)$$

\noindent with\ R\ the radius of the heavy onium.

Eq.(5) breaks down at large \ Y\ for two separate reasons.  (i)  Even though the heavy 
oniun
state has a small radius, \ R,\ the gluons responsible for the growth of (3) and (4) with
increasing \ Y\ diffuse to larger distances, determined by the final exponential factors in 
(3)
and (4).  When the diffusion reaches momenta as small as the QCD $\wedge$-parameter 
the whole
perturbative approach breaks down.  (ii)  When \ Y\  becomes large the cross section 
clearly
grows faster than unitarity allows so that new corrections must become important which 
modify 
simple BFKL behavior.

Consider first the question of diffusion.  Eqs.3 and 4 have been derived neglecting the 
running
of the QCD coupling.  Let $k_0=1/R.$  Then the minimum momentum to which gluons 
diffuse in an
onium-onium collision in $k_m$ given by

$$\ell n^2 k_o/k_m = D Y\eqno(6)$$

\noindent as determined from (4) taking $[exp\{-{\ell n^2k_0/k_m\over 4 D\  Y/2}\}]^2 = 
1/e$
because diffusion is maximum at rapidities midway between the two onia.  (We shall give 
a more
complete derivation of (6) later on.)  Then the condition for the running of the coupling to 
be
negligible over the region where gluon momenta diffuse is

$${\alpha(k_m) - \alpha(k_0)\over \alpha(k_m)} << 1\eqno(7)$$

\noindent which leads to

$$Y << {\pi\over 14 N_c\zeta(3) b^2}\ \ {1\over \alpha^3(k_0)}\eqno(8)$$

\noindent with $b = {11 N_c-2N_f\over 12\pi}$ and where we have used $\alpha(k) = 
{1\over b\ell n\ 
k^2/\wedge^2}.$

\noindent If one is willing to use a running coupling in the BFKL equation, rapidity 
values

$$Y \leq {\pi\over 14 N_c\zeta(3) b^2}\ \ {1\over \alpha^3(k_0)} \eqno(9)$$

\noindent can be reached.  One cannot go beyond the limit set in (9) without having 
$k_m$-values
less than $\wedge.$

The limit (8), or (9), is not too bad because unitarity constraints are expected to be 
reached,
from (5), when rapidities become of size

$$Y \leq {2\over \alpha_P-1} \ \ell n {1\over \alpha(k_0)}.\eqno(10)$$

\noindent For $\alpha(k_0)$ very small unitarity constraints become important long 
before
diffusion constraints.  Thus questions of unitarity can in principle be studied completely 
within
the fixed coupling approximation[15] for sufficiently heavy onia.  However, when 
$\alpha(k_0)$ is
of moderate size the diffusion limit (9) may be reached before unitarity corrections are
important.  In our leading logarithm discussion the limits (9) and (10) are comparable at
$\alpha(k_0)$ is of size 1/3 to 1/2  although numerical studies suggest that one in fact
must go significantly beyond (10) to see strong unitarity corrections[15].  Thus in most
$practical$ circumstances diffusion constraints will be reached before unitarity 
constraints.

\section{Incorporating BFKL evolution in a DGLAP formalism}

The asymptotic behavior (4) follows directly from the representation[4,5,16]

$$T(Y, Q/\mu) = \int {d\lambda\over 2\pi i} exp\{{2\alpha N_c\over \pi} \chi(\lambda)Y 
+ \lambda
\ell n\  Q^2/\mu^2\}\eqno(11)$$

\noindent where

$$T(Y, Q/\mu) = {\pi Q^2\over 4\alpha^2} f(Q,Y,\mu),\eqno(12)$$

\noindent with

$$\chi(\lambda) = \psi(1) - {1\over 2} \psi(\lambda) - {1\over 2} \psi(1-
\lambda)\eqno(13)$$

\noindent and where the path of integration in (11) goes from $\lambda = \lambda_0-i\  
\infty$ to
$\lambda_0 + i\  \infty$ with $0 < \lambda_0 < 1.$  In order to cast this into a form\ 
corresponding to the operator product expansion it is necessary to pick out the leading 
twist part
of (11).  If we take $Q/\mu > 1$ then higher powers of $(\mu/Q)^2$ can discarded simply 
by
changing  the
$\lambda$-integration in (11) to a circle, of radius less than 1, about the origin.  Call this
contour  C.  Then the leading twist part of  T, denoted by  \ L,\ is

$$L(Y,Q/\mu)\  \equiv \int_C{d\lambda\over 2\pi i} exp\{{2\alpha N_c\over \pi} 
\chi(\lambda)Y +
\lambda \ell n\  Q^2/\mu^2\}$$
$$= \int {dn\over 2\pi i} C_n exp\{\gamma_n\ell n\  Q^2/\mu^2 +
(n-1)Y\}\eqno(14)$$

\noindent  and determines the anomalous dimension and coefficient functions[17].  The 
n-integration
runs parallel to the imaginary axis and to the right of the point n=1.  (The anomalous 
dimension,
$\gamma_n,$ is unique in the leading logarithmic BFKL approximation.  The coefficient 
function,
$C_n,$ is not unique but the choice above, where matrix elements are normalized to 1 at 
$\mu$, is
natural and convenient for our purposes.)  We note that when \ Y\ is large subleading 
twist terms
are also subleading in  \ Y\ since $\lambda = 1/2$ is the largest of all the saddle points of
$\chi(\lambda).$

Using (14) it is straightforward to find[16]

$$\gamma_n=\chi^{-1}({\pi(n-1)\over 2\alpha N_c})\eqno(15)$$

\noindent and

$$C_n= - {d\gamma_n\over dn}\eqno(16)$$

\noindent where $\chi^{-1}$ is the function inverse to $\chi.$  If one writes

$$\gamma_n = \sum_{N=1}^\infty \gamma^{(N)}({\alpha N_c\over \pi(n-
1)})^N\eqno(17)$$

\noindent and

$$C_n=\sum_{N=0}^\infty C^{(N)}({\alpha N_c\over \pi(n-1)})^N\ {1\over n-
1},\eqno(18)$$

\noindent then (16) gives $C^{(N)} = N\gamma^{(N)}.$  Using (14) one finds

$$C^{(N)} = N! \sum_{k=0}^{N-1}\ {1\over k!(k+1)!\ (N-1-k)!}\ {d^k\over
d\lambda^k}[\rho(\lambda)]^{N-1-k}\bigg\vert_{\lambda=0}\eqno(19)$$

\noindent with $C^{(0)}=1$ and

$$\rho(\lambda) = 2\chi(\lambda) - {1\over \lambda}.\eqno(20)$$

\noindent (Eq.19 is most easily found by setting $Q=\mu$ in (14), and then taking the 
term $Y^N$
on each side of that equation.)

The coefficient function and anomalous dimension have a singularity at $n=\alpha_P$ 
corresponding
to the saddle point of $\chi(\lambda)$ at $\lambda = 1/2.$  One can determine the 
behavior of
$C_n$ and $\gamma_n$ near $n=\alpha_P$ either from (15) and (16) using

$$\chi(\lambda) = 2 \ell n 2 + 7 \zeta(3)(\lambda -{1\over2})^2 + \cdot \cdot 
\cdot\eqno(21)$$

\noindent or from (17) and (18) using

$$C^{(N)} 
\begin{array}{c}
{ }\\{\sim}\\^{N\to \infty}
\end{array}
\ {(4\ell n\  2)^N\over \sqrt{N}}\ \sqrt{{\ell 
n\ 
2\over 14\pi
\zeta(3)}}.
\ \ \ \ \ \ \ \ \ \ \ \ \ \ \ \ \ \ \ \ \ \ \ \ \ \ \ \ \ \ \ \ \ 
\ (22)$$

\noindent Near $n = \alpha_P$ one finds[4,5,6-8,17]

$$C_N \approx \   {1\over 4\sqrt{ D{(n-\alpha_P)}}}.\eqno(23)$$

\noindent and

$$\gamma_n-{1\over 2} \approx - {1\over 2} {\sqrt{{(n-\alpha_P)\over 
D}}}.\eqno(24)$$

We are now in a position to see more clearly how diffusion contributes to L.  It is 
convenient to
separate the non-diffusion parts of \ L\ according to

$$L\ =\ {1\over 2} \tilde{L}(Q/\mu)\ e^{(\alpha_P-1)Y}\eqno(25)$$

\noindent where we have normalized $\tilde{L}$ to be a probability distribution.  Then,

$$\tilde{L}(Y, Q/\mu) = \ {1\over 2} \int {dn\over 2\pi{i}}\ C_n exp\{(\gamma_n-
{1\over 2}) \ell
n\  Q^2/\mu^2\ + (n-\alpha_P)Y\}\eqno(26)$$

\noindent obeys, at large\ Y,\ the diffusion equation[4,5,13]

$$\left({\partial\over \partial Y} - D {\partial^2\over \partial \ell n^2\ 
Q/\mu}\right)\tilde{L} = O,
\eqno(27)$$

\noindent as is easily seen using (24) in (26).  The solution is

$$\tilde{L}(Y, Q/\mu) = {exp\{-{\ell n^2\  Q/\mu\over 4DY}\}\over \sqrt{4\pi 
DY}}\eqno(28)$$

\noindent which, using (25) and (12), gives (4),

\section{Limitations on the use of DGLAP evolution at small x}

Our normal picture of DGLAP evolution is not one of diffusion but of a monotonic 
increase of
$\ell n\  Q/\mu$ with Y.  In order to see where the diffusion is hidden write\ L,\ given in 
(14),
as

$$L(Y, Q/\mu) = \int_0^Y dy A(y, Q/\mu) C(Y-y)\eqno(29)$$

\noindent where asymptotically,

$$C(Y) = \int {dn\over 2\pi{i}} C_n e^{(n-1)Y} \sim {e^{(\alpha_P-1)Y}\over 
\sqrt{16\pi DY}}$$

\noindent and

$$A(Y,Q/\mu) = \int {dn\over 2\pi{i}} e^{\gamma_n \ell n\  Q^2/\mu^2 + (n-1)Y}$$
\ \ \ \ \ \ \ \ \ \ \ \ \ \ \ \ $$\sim {Q\  \ell n
Q^2/\mu^2\over \mu\  Y}\ {e^{(\alpha_P-1)Y}\over \sqrt{16\pi DY}} exp\{-{\ell n^2\  
Q/\mu\over 4
DY}\}\eqno(30)$$

\noindent A straightforward calculation shows that the values of  y  in (29) which 
dominate the
integral are

$$y \propto {Y\over 1 + ({4 D Y\over \ell n^2\  Q/\mu})}.$$

\noindent Thus, for $Y >> {\ell n^2\  Q/\mu\over 4 D}$ the coefficient function covers 
most of the
rapidity region in \ L,\ as given in (29), so that for very large values of \ Y\ most of the
diffusion is hidden in C.

However, this is not the whole story.  There is also significant diffusion in\ A\ despite
the  fact that the  DGLAP equation

$${\partial A(Y, Q/\mu)\over \partial \ell n\  Q/\mu} = \int_0^Y dy\  \gamma(y)
A(Y-y,Q/\mu).\eqno(31)$$

\noindent seems to suggest increasing Q-values and increasing Y-values go together.  To 
see where
the diffusion is hidden it is convenient to write

$$A = {Q\over \mu} e^{(\alpha_P-1)Y} \tilde{A}(Y, Q/\mu)\eqno(32)$$

\noindent So that $\tilde{A}$ obeys

$${\partial\tilde{A}\over \partial \ell n\  Q/\mu} = \int_0^Y dy(\gamma - 
1/2)(y)\tilde{A}(Y-y,
Q/\mu)\eqno(23)$$

\noindent with

$$(\gamma - {1\over 2})(Y) = \int {dn\over 2\pi{i}} (\gamma_n - {1\over 2})
e^{(n-\alpha_P)Y}.\eqno(34)$$

\noindent It is now straightforward to see that the y-values dominating (33) are given by

$$y\  \alpha\  {4 D Y\over \ell n^2\  Q/\mu + 4 D Y} Y. \eqno(35)$$

\noindent Thus when $Y \geq {\ell n^2\  Q/\mu\over 4 D}$ the action of the anomalous 
dimension is
very nonlocal in the DGLAP equation (33).  This allows the diffusion, which according 
to (29)
is certainly contained in A, to be hidden in the  anomalous dimension  $\gamma.$

To see the general limitations on what values of \ Y\ allow a consistent operator product
expansion, and a DGLAP formalism, we note that

$$L(Y, Q/\mu) = \int_0^\infty {dk^2\over k^2} L(Y-y, Q/k) L(y, k/\mu).\eqno(36)$$

\noindent From (36) one determines , using (25) and (28), that the smallest important 
values of \
k\ in (36) are reached when one takes

$$y=y_m = {Y\over 2} (1-{\ell n\  Q/\mu\over \sqrt{4DY}})\eqno(37)$$

\noindent with the effective minimum value, $k_m,$ given by

$$\ell n^2\  {Q\mu\over k_m^2} = 4DY.\eqno(38)$$

\noindent Taking $k_m=\wedge$ as the mimimum allowed value for $k_m$ before the 
whole formalism
breaks down we get the restriction

$$Y \leq {1\over 4D} \ell n^2\  {Q\mu\over \wedge^2} = {1\over 4D}(\ell n\  Q/\mu + 
\ell n\ 
\mu^2/\wedge^2)^2.\eqno(39)$$

\noindent If $\ell n\  \mu^2/\wedge^2 << \ell n\  Q/\mu$ one finds $Y \leq {1\over 4D} 
\ell n^2\ 
Q/\mu$ and the breakdown occurs because of diffusion in the high order corrections to 
the
anomalous dimension.  If $\ell n\  Q/\mu << \ell n\  \mu^2/\wedge^2$ one finds $Y \leq 
{1\over 4D}
\ell n^2\ 
\mu^2/\wedge^2$ and the breakdown occurs in the coefficient function or, in more 
general schemes, in a combination
of the coefficient function and the operator matrix element.  In applying (39) to deep 
inelastic
scattering one should identify $\mu$ with the transverse momentum scale where 
perturbation starts
to apply while $Y = \ell n\  \bar{x}/x$ where $\bar{x}$ as the x-value where soft gluon 
emission
starts to become important.  As a rough estimate we take $\mu = \wedge$ and we suppose 
$\bar{x}
\approx 1/10.$  Then

$$\ell n\  \bar{x}/x \leq {1\over 4D} \ell n^2 Q/\wedge\eqno(40)$$

\noindent gives the region of \ x\ where the operator product expansion applies.  (At our 
level 
of discussion we are unable to fix the scale at which $\alpha$ should be\ evaluated in 
D.  $\alpha = \alpha(Q)$ would certainly give an upper bound in (40).)

If one wishes to use the DGLAP equation to evolve from $Q_0$ to $Q$ one must satisfy 
(39) with
$\mu = Q_0.$  However, the stronger constraint is that the operator product expansion be 
valid at
$Q_0$, that is one needs $\ell n\  \bar{x}/x \leq {1\over 4D} \ell n^2\  Q_0/\wedge$ for 
x-values
which are important at $Q_0$ in determining the parton distribution at the desired small 
value of
x.  In case one chooses $Q_0 \geq Q$ then the strongest constraint is (40), the 
requirement that
the operator product expansion apply at the \ x\ and $Q^2$ value one is interested in.

At the level we have calculated no breakdown of the operator product expansion is yet
visible[10,11].  However, once momenta as low as $\wedge$ occur,and that will happen 
when (39) is
not satisfied, higher order running coupling corrections will not be small.  This signifies 
the
fact that the anomalous dimension and/or the coefficient function cannot be calculated
perturbatively.  Since this breakdown of perturbative occurs at the leading twist level it
corresponds to the breakdown of the operator product expansion itself.  Also, since our
discussion concerns leading twist effects this phenomena is different from the appearance 
of
renomalons.  The breakdown of the perturbation series when higher order running 
coupling effects
are included will here not be associated with an n!  growth of the perturbation series, but
simply with the fact that higher order running coupling corrections are not small.  It is 
likely
that diffusion puts significant restrictions on the use of a DGLAP analysis of present 
small-x
and small-$Q^2$ deep inelastic scattering data.  At the same time we are presented with
the interesting challenge of  how to analyze moderate $Q^2$ and very low x phenomena.

\end{document}